\documentclass[a4paper,10pt]{llncs}

\usepackage[colorlinks=true,urlcolor=blue,citecolor=black,linkcolor=red,final]{hyperref}
\usepackage[utf8]{inputenc}
\usepackage[english]{babel}
\usepackage{pdfcomment}
\usepackage[final]{pdfpages}
\usepackage[T1]{fontenc}
\usepackage{amssymb, xspace,multicol}
\usepackage{listings,amsmath,dsfont}
\usepackage{sectsty}
\usepackage{ifthen}
\usepackage{url}
\usepackage{caption}
\usepackage{framed}
\title{Polynomial invariants by linear algebra}
\author{Steven de Oliveira$^1$, Saddek Bensalem$^2$, Virgile Prevosto$^1$}
\institute{CEA, LIST, Software Reliability and Security Lab\\
\email{\{steven.deoliveira,virgile.prevosto\}@cea.fr}
\and
Université Grenoble Alpes\\
\email{saddek.bensalem@imag.fr}}

\newcommand{\toolname}{\textsc{Pilat}\xspace}
\newcommand{\tooldef}{{\bf P}olynomial {\bf I}nvariants by {\bf L}inear 
{\bf A}lgebra {\bf T}ool}

\newcommand{\aligator}{\textsc{Aligator}\xspace}
\newcommand{\fastind}{\textsc{Fastind}\xspace}

\newcounter{thm}
\newcounter{prop}

\newtheorem{Def}{Definition}

\renewcommand{\k}{\ensuremath{\mathbb{K}}}
\newcommand{\kbar}{\ensuremath{\overline{\k}}}
\newcommand{\1}{ \ensuremath{\mathds{1}}}
\newcommand{\scalprod}[2]{\ensuremath{\left<#1,#2\right>}}
\newcommand{\vect}{\ensuremath{\mathrm{Vect}}}

\newcommand{\longarticle}{1}

\newcommand{\annexedProof}{1}

\newcommand{\pathtoprops}{}

\newcommand{\inAnnex}{0}

\newcommand{\labelprop}[1]{
  \ifthenelse{\inAnnex = 0}
  {\label{#1}}
  {} 

}

\newcommand{\propinput}[1]{
  \ifthenelse{\longarticle = 0}{\input{\pathtoprops#1}} {
  \ifthenelse{\annexedProof = 0}
    {\ifthenelse{\inAnnex = 0}
      {\input{\pathtoprops#1} \input{\pathtoprops#1_proof}}
      {}
    }
    {\ifthenelse{\inAnnex = 0}
      {\input{\pathtoprops#1}}
      {\input{\pathtoprops#1} \input{\pathtoprops#1_proof}}
    }
  }
  }
  
\lstdefinelanguage{pilat}{
morekeywords={OR,while,do,done},
}
\lstdefinestyle{pilatstyle}{
language=pilat,
basicstyle={},
identifierstyle={\itshape},
keywordstyle={\bfseries},
mathescape=true
}

\begin{document}

\maketitle

\begin{abstract}
      We present in this paper a new technique for generating polynomial invariants,
    divided in two independent parts : a 
    procedure that reduces polynomial assignments composed loops analysis to
    linear loops under certain hypotheses and a procedure for generating 
    inductive invariants for linear loops.
    Both of these techniques have a polynomial complexity for a bounded number 
    of variables and we guarantee the completeness of the
    technique for a bounded degree which we successfully implemented for 
    C programs verification.

\end{abstract}

  \section{Introduction} 
   
  When dealing with computer programming, anyone should be aware of the
  underlying behavior of the whole code, especially when it comes to 
  life-critical projects composed of million of
  lines of code~\cite{hoare2003verifying}.
  Manual code review cannot scale to the size of actual embedded programs.
  Testing allows to detect many vulnerabilities but it is never enough to 
  certify their total absence. Indeed, the cost of generating and executing 
  sufficient test cases to meet the most stringent coverage 
  criteria~\cite{botella2009automating} that are expected for critical
  software becomes quickly prohibitive as the size of the code under test 
  grows. 
Alternatively, formal methods techniques based on abstraction allow us to prove 
the absence of error. 

  However, since a program can, at least in theory,
  have an infinite number of different 
  behaviors, the verification problem is undecidable and these 
  techniques either lose precision (emitting false alarms) and/or require 
  manual input.
  One of the main issue of such approach is the analysis of loops, considered
  as a major research problem since the 70s~\cite{basu1975proving}.
  Program verification based on Floyd-Hoare's inductive 
  assertion~\cite{hoare1969axiomatic} and CEGAR-like 
  techniques~\cite{clarke2000counterexample} for model-checking
  uses loop invariants in order to reduce the problem to an acyclic graph 
  analysis~\cite{beyer2007path} instead of unrolling or accelerating 
  loops~\cite{hojjat2012accelerating}.
  Thus, a lot or research nowadays is focused on the 
  automatic inference of loop
  invariants~\cite{kovacs2010complete,rodriguez2007generating}.

  We present in this paper a new technique for generating polynomial 
  invariants, divided in two independent parts : a 
  \emph{linearization} procedure that reduces the analysis of solvable loops, 
  defined in~\cite{rodriguez2007generating}, to the analysis of linear loops ;
  an \emph{inductive invariant generation} procedure for linear loops.
  Those two techniques are totally independent from each other, we aim to
  present in this article their composition in order to find polynomial 
  invariants for polynomial loops.
  We also add an extension of this composition allowing to treat loops with
  complex behaviors that induces the presence of complex
  numbers in our calculation.
  The linearization algorithm has been inspired by a compiler optimisation 
  technique called \emph{operator strength reduction}~\cite{cooper2001operator}.  
  Our invariant generation is completely independent from the initial
  state of the loop studied and outputs parametrized invariants, which is very
  effective on programs using a loop multiple times and loops for which we have  
  no knowledge of the initial state.
  In addition to being complete for a certain class of polynomial relations, 
  the invariant generation technique has the 
  advantage to be faster than the already existing one for such loops 
  as it relies on polynomial complexity linear algebra algorithms.

  Furthermore, a tool implementing this method has been developped 
  in the Frama-C framework for C programs verification~\cite{Kirchner2015} as
  a new plug-in called \toolname (standing for \tooldef).
  We then compared our performances with \aligator~\cite{kovacs2010complete}
  and \fastind~\cite{cachera2014inference}, two 
  invariant generators working on similar kinds of loops. 
  First experiments over a representative
  benchmark exposed great improvements in term of computation time.
  
  \paragraph{Outline.}
 The rest of this paper is structured as follows.
  Section~\ref{prel} introduces the theoretical concepts used all along the
  article and the kind of programs we focus on.
  Section~\ref{overview} presents the application of our technique on a 
  simple example.
  Section~\ref{poly2lin} presents the linearization step for simplifying
  the loop, reducing the problem to the study of affine loops. 
  Section~\ref{inv_gen} presents our contribution for generating all 
  polynomial invariants of affine loops.
  Section~\ref{complex} extends the method with the treatment of invariants
  containing non-rational expressions.
%
  Finally, section~\ref{results} compares \toolname to \aligator and \fastind.
  \ifthenelse{\longarticle = 1}
  {}
  {Due to space constraints, proofs have been omitted. They are available 
  in a separate report~\cite{pilat_long}}.
 
  \paragraph{State of the art.}
 
  Several methods have been proposed to generate invariants for kinds of loops
  that are similar to the ones we address in this paper. In particular,
  the weakest precondition calculus of polynomial
  properties in~\cite{muller2004precise} is based on the computation of 
  the affine transformation kernel done by the program.
  This method is based on the computation of the kernel of the affine
  transformation described by the program.
      More than requiring the whole program to be affine, 
      this method relies on the
      fact that once in the program there exists a non-invertible
      assignment, otherwise the kernel is empty.
      This assumption is valuable in practice, 
      as a constant initialization is non-
      invertible, so the results may appear at the end of a whole-program 
      analysis and highly depend on the initial state of the program. 
      On the other hand,
      our method can generate parametrized invariants, computable
      without any knowledge of the initial state of a loop, making it more 
      amenable to modular verification.
    
    From a constant propagation technique 
    in~\cite{muller2002polynomial} to a complete invariant generation 
    in~\cite{rodriguez2007generating}, Gröbner bases have proven to be 
    an effective way to handle polynomial invariant generation. 
    Such approaches have been successfully implemented in the tool 
    \aligator~\cite{kovacs2010complete}. 
    This tool generates all polynomial invariants of any degree from a succession of 
    $p$-solvable polynomial mappings in very few steps. 
    It relies on the iterative computation of Gröbner bases of some polynomial
    ideals, which is a complicated problem proven to be
    EXPSPACE-complete~\cite{mayr1989membership}.
    
    Attempts to get rid of Gröbner bases as 
    in~\cite{cachera2014inference} using abstract interpretation
    with a constant-based instead of a iterative-based technique
    accelerates the computation of invariants by generating 
    abstract loop invariants. 
    However, this technique is incomplete and misses some invariants. The
    method we propose here is complete for a particular set of loops defined 
    in~\cite{rodriguez2007generating} in the sense that it finds all
    polynomial relations $P$ of a given degree verifying $P(X) = 0$ at every 
    step of the loop, and has a polynomial complexity
    in the degree of the invariants seeked for a given number of variables.

  \section{Preliminaries}\label{prel}
  
\paragraph{Mathematical background.}
  Given a field $\k$, $\k^n$ is the vector space of dimension $n$ 
  composed by vectors with $n$ coefficients in $\k$. 
  Given a family of vector $\Phi \subset \k^n$, $Vect(\Phi)$ is the vector
  space generated by $\Phi$.
  Elements of $\k^n$ are denoted $x = (x_1,...,x_n)^t$ a column vector.
  $\mathcal{M}_n(\k)$ is the set of 
  matrices of size $n*n$ and $\k[X]$ is the set of polynomials
  using variables with coefficients in $\k$. 
  We note $\kbar$ the algebraic closure
  of $\k$, $\kbar = \{x | \exists P \in \k[X], P(x) = 0\}$.
  We will use $\scalprod{.}{.}$ the linear algebra standard notation, $\scalprod{x}{y} = x\cdot y^t$,
  with $\cdot$ the standard dot product. 
  The kernel of a matrix $A \in \mathcal{M}_n(\k)$, denoted $\ker(A)$, is the 
  vector space defined as $\ker(A) = \left\{ x | x \in \k^n, A.x = 0\right\}$.
  Every matrix of $\mathcal{M}_n(\k)$ admits a finite set of eigenvalues 
  $\lambda \in \kbar$ and their associated eigenspaces $E_\lambda$, defined as
  $E_\lambda = \ker(A - \lambda Id)$, where $Id$ is the identity matrix and 
  $E_\lambda \neq \{ 0\}$. 
  Let $E$ be a $\k$ vector space, $F \subset E$ a sub vector space of $E$ 
  and $x$ an element of $F$.
  A vector $y$ is \emph{orthogonal} to $x$ if $\scalprod{x}{y} = 0$. We denote $F^\perp$ 
  the set of vectors orthogonal to every element of $F$.

\paragraph{Programming model.}

  We use a basic programming language whose syntax is given in 
  figure~\ref{fig:code_semantics}. $Var$ is a set of variables that can
  be used by a program, and which is supposed to have a total order.
  Variables take value in a field $\mathbb{K}$. A program state
  is then a partial mapping $Var \rightharpoonup \k$. 
  Any given program only uses a finite number $n$ of variables.
  Thus, program states can be represented as a
  vector $X = (x_1,...,x_n)^t$. In addition, we will note 
  $X' = (x'_1,...,x'_n)^t$ the program state after an assignment. 
  Finally, we assume that for all programs, there
  exists $x_{n+1} = x'_{n+1} = \1$ a constant variable always equal to $1$.
  \begin{figure*}[thbp]
    \begin{framed}
     
    \begin{center}
     
    \begin{multicols}{2}
    
    \begin{tabular}{rcl}
      $i$ & ::= &  \lstinline|i;i| \\
          & | & \lstinline|($x_1$,..,$x_n$):=($exp_1$,...,$exp_n$)| \\
          & | & \lstinline|i OR i| \\
          & | & \lstinline|while (*) do i done|
    \end{tabular}
    \columnbreak 
    
      \begin{tabular}{lcl}
      $exp$ ::= && \xspace\xspace $cst \in \k$\\
      \xspace &&| $x \in Var$\\
      \xspace &&| $exp + exp$ \\
      \xspace &&| $exp * exp$ \\
    \end{tabular}
    
    \end{multicols}
    \end{center}
    
    \end{framed}

    \caption{\label{fig:code_semantics} Code syntax}
\end{figure*}

The $i$ OR $i$ instruction refers to a non-deterministic condition.

Each $i$ will be refered to as one of the \emph{bodies} of the loop.

Multiple variables assignments occur simultaneously within a single instruction.
We say that an instruction is affine when it is an assignment 
for which the right values are affine. 
If not, we divide instructions in two 
categories with respect to the following definition, 
from~\cite{rodriguez2007generating}

 \begin{Def}\label{solv_map}
 Let $g\in\mathbb{Q}[X]^m$ be a polynomial mapping. $g$ is solvable if there exists 
 a partition of $X$ into subvectors of variables $x = w_1\uplus ...\uplus w_k $
 such that $\forall j,~1\leqslant j \leqslant k$ we have 
 \begin{equation*}
  g_{w_j}(x) = M_jw_j^T + P_j(w_1,...,w_{j-1})
 \end{equation*}
 
 with $(M_i)_{1 \leqslant i \leqslant k}$ a matrix family
 and $(P_i)_{1 \leqslant i \leqslant k}$ a family of polynomial mapping.
 
 \end{Def}
 
An instruction is solvable if the associated assignment is a 
solvable polynomial mapping. Otherwise, it is unsolvable.
Our technique focuses on loops containing only solvable instructions,
thus it is not possible to generate invariants for nested loops.
It is however possible to find an invariant for a loop containing no inner loop
even if it is itself inside a loop, that's why we allow the construction.

  \section{Overview of our approach}\label{overview}
  \paragraph{Steps of the generation.}
In order to explain our method we will take the following running example, 
for which we want to compute all invariants of degree $3$: 

\begin{lstlisting}
 while (*) do
    (x,y) := (x + y*y, y + 1)
 done
\end{lstlisting}
Our method is based on two distinct parts : 
\begin{enumerate}
 \item reduction of the polynomial loop to a linear
 loop;
 \item linear invariant generation from the linearized loop.
\end{enumerate}

We want to find a linear mapping $f$ that \emph{simulates} the behavior of 
the polynomial mapping $P(x,y) = (x + y^2, y + 1)$. 
To achieve this, we will express the value of every monomial
of degree 2 or more using brand new variables.
Here, the problem comes from the $y^2$ monomial.
In~\cite{muller2004precise}, it is described how
to consider the evolution of higher degree monomials as affine applications of lower 
or equal degree monomials when the variables involved in those monomials evolve 
affinely. 
We extend this method to express monomials transformations of the loop by 
affine transformations, reducing the problem to a simpler loop analysis.
For example here, $y' = y+1$ is an affine assignment, so there exists an 
affine representation of $y_2 = y^2$, which is $y_2' = y_2 + 2.y + 1$. 
Assuming the initial $y_2$ is correct, we are sure to express the value of 
$y^2$ with the variable $y_2$. 
Also, if we want to find invariants of degree 
$3$, we will need to express all monomials of degree $3$,
i.e. $xy$ and $y_3$ the same way. (monomials containing $x^i$ with 
$i \geqslant 2$ are irrelevant as their expression require the expression
of degree $4$ monomials).
Applying this method to $P$ gives us the linear mapping 
$f(x,y,y_2,xy,y_3, \1) = (x + y_2, y +  \1, y_2 + 2.y +  \1,xy + x + y_2 + y_3,y_3 + 3.y_2 + 3.y +  \1,\1)$, 
with $\1$ the constant variable mentioned in the previous section.

Now comes the second part of the algorithm, the invariant generation.
Informally, an invariant for a loop is a formula that
\begin{enumerate}
 \item is valid at the beginning of the loop ;
 \item stays valid after every loop step.
\end{enumerate}

We are interested in finding \emph{semi-invariants} complying only with
the second criterion such that they can be expressed as a linear equation over
X, containing the assignment's original variables and the new ones
generated by the linearization procedure.
In this setting, a formula satisfying the second criterion
is then a vector of coefficients $\varphi$ such that 
\begin{equation}\label{semi_inv}
 \scalprod{\varphi}{X} = 0 \Rightarrow \scalprod{\varphi}{f(X)} = 0
\end{equation}
By linear algebra, the following is always true 
\begin{equation}\label{eq_dual}
\scalprod{\varphi}{f(X)} = \scalprod{f^*(\varphi)}{X}
\end{equation}
where $f^*$ is the dual of $f$.
If $\varphi$ happens to be an eigenvector of $f^*$ (i.e. there exists $\lambda$
such that $f^*(\varphi) = \lambda \varphi)$,
the equation~(\ref{semi_inv})~becomes 
\begin{eqnarray*}
\scalprod{\varphi}{X} = 0 &\Rightarrow& \scalprod{f^*(\varphi)}{X} = 0\mathrm{~by~(\ref{eq_dual})}\\
\scalprod{\varphi}{X} = 0 &\Rightarrow& \scalprod{\lambda.\varphi}{X} = 0\\
\scalprod{\varphi}{X} = 0 &\Rightarrow& \lambda.\scalprod{\varphi}{X} = 0
\end{eqnarray*}
which is always true.
We just need to \emph{transpose} the matrix 
representing $f$ to compute $f^*$ . It returns 
$f^*(x,y,y_2,y_3,\1) = (x,y + y_2 + y_3, x + y_2 + 3.y_3,y_3,y + y_2 + y_3 + \1,y + y_2 + y_3 + \1)$.
$f^*$ only admits the eigenvalue $1$. The eigenspace of $f^*$ associated to $1$ 
is generated by two independants vectors, $e_1 = (-6,1,-3,2,0)^t$ and
$e_2 = (0,0,0,0,1)^t$.
Eventually, we get the formula $F_{k1,k2} = (k_1.(-6.x + y -3.y_2 + 2.y_3) + k_2.\1 = 0)$ as
invariant, with $k_1,k_2 \in \mathbb{Q}$. By writing $k = -\frac{k_2}{k_1}$ and
replacing $\1$ with $1$, we can rewrite it with only one parameter,
$F_k = (-6.x + y -3.y_2 + 2y_3 = k)$.
In this case, information on the initial state of the loop allows to fix
the value of the parameter $k$.
For example if the loop starts with $(x = 0, y = 0)$, 
then $-6.x + y -3.y^2 + 2.y^3 = 0$, and $F_0$ is an invariant.
The next section will show how the work done on our example can be generalized
on any (solvable) loop. In particular, section~\ref{poly2lin} will deal with the
linearization of polynomial assignments. 
Then we will see in section~\ref{inv_gen} that the eigenspace of 
the application actually represents all the possible invariants of $f$ and that
we can always reduce them to find a formula with only one parameter.
 
\paragraph{Extension of the basic method.}

The application's eigenvector may not always be rational. 
For example, applying the previous technique on a mapping such as 
$f(x,y) = (y, 2.x)$ will give us invariants with coefficients 
involving $\sqrt2$.
Dealing with irrational and/or complex values raises some issues in our current 
implementation setting. 
Therefore, we propose in section~\ref{complex} a solution to stick with 
rational numbers.
Eventually, we treat the case when a 
condition occur in loops in section~\ref{mult_loop}.

 \section{Automated generation of loop invariants}
  
  \subsection{Strength reduction of polynomial loops}\label{poly2lin}
  
\paragraph{Lowerization.}

%
  Let $P$ be a program containing a single loop with a single solvable 
  assignment
  \lstinline|X := g(X)|.   
  In order to reduce the invariant generation problem for solvable polynomial
  loops to the one for affine loops, we need to find a linear mapping $f$ 
  that perfectly matches $g$.
\begin{figure*}[htbp]
 
\begin{framed}
L1 :  
\begin{lstlisting}
while (*) do	
 (x,y,z) := (x + 1, y + 2, z + x*y)
done
\end{lstlisting}
 
L2 :
\begin{lstlisting}
xy = x*y
while (*) do
  (x,y,xy,z) := (x + 1, y + 2, xy + 2x + y + 2, z + xy)
done
\end{lstlisting}
   
\end{framed}

 \caption{\label{fig:loop_ex} Polynomial and affine loop having the same behavior}
  \end{figure*}
As shown in figure~\ref{fig:loop_ex}, the first loop L1 is polynomial but there 
exists a similar affine loop, namely L2, 
computing the same vector of values plus and thanks to an extra variable $xy$.

 \begin{Def}\label{def:linearizable}
  Let $g$ be a polynomial mapping of degree $d$ using $m$ variables. $g$ is 
  linearizable if there exists a linear mapping 
  $f$ such that $X' = g(X) \Rightarrow (X',P(X')) = f(X,P(X))$, 
  where $P : \mathbb{Q}^m \rightarrow \mathbb{Q}^n$ is a polynomial of degree 
  $d$.
 \end{Def}

 
 By considering polynomials as entries of the application, we are able to 
 consider the evolution of the polynomial value instead of recomputing it for 
 every loop step. 
 This is the case in the previous example, where the 
 computation of $xy$ as $x * y$ is made once at the beginning of the loop.
 Afterwards, its evolution depends linearly of itself, $x$ and $y$. 
 Similarly, if we want to consider $y^n$ for some $n\geq 2$, 
we would just need to 
 express the evolution of $y^n$ by a linear combination of itself and 
 \emph{lower degree monomials}, which could themselves be expressed as linear 
 combinations of lower degree monomials, until we reach an affine application.
  We call this process the polynomial mappings \emph{lowerization} or 
  \emph{linearization}.
 
 \paragraph{Remark.}
 This example and our running example have the good property to be 
 linearizable. However, this property is not true for all polynomials loops. 
 Consider for example the mapping $f(x) = x^2$. Trying to express $x^2$ as a 
 linear variable will force us to consider the monomials $x^4$, $x^8$ and so 
 on. Thus, we need to restrain our study to mappings that 
 \emph{do not polynomially transform a variable itself}. 
  This class of polynomials corresponds to solvable polynomial mappings, 
  defined in Definition~\ref{solv_map}.

\propinput{linearizable}

  For example, let $g(x,y) = (x + y^2,y+1)$.
  $g$ is linearized by $f(x,y,y_2) = (x+y_2,y+1,y_2 + 2y + 1)$.
  Indeed with $(x',y') = g(x,y)$, we have $(x',y',y'^2 ) = f(x,y,y^2)$
 
%
%
%

 \paragraph{Linearization Algorithm.}
 
 The algorithm is divided in two parts : the solvability verification of 
 the mapping and, if successful, the linearization process.
 The solvability verification consists in finding an appropriate partitioning 
 of the variables that respects the solvable constraint. 
 It is nothing more than checking that a variable $v$ cannot be in
 a polynomial (i.e. non linear) assignment of another variable that itself 
 depend on $v$.
 This check can be reduced to verifying the acyclicity of a graph, which can 
 be computed e.g. by Tarjan's~\cite{tarjan1972depth} or 
 Johnson's~\cite{johnson1975finding} algorithms.

 The linearization process then consists in considering all monomials
 as new variables, then finding their linear evolution by replacing
 each of their variables by the transformation made by the initial
 application. 
 This may create new monomials, for which we similarly create new variables
 until all necessary monomials have been mapped to a variable.
 Since we tested the solvability of the loop, the variable creation process 
 will eventually stop. Indeed, if this was not the case, this would mean that 
 a variable $x$ transitively depends on $x^d$ with $d > 1$. 
 
\paragraph{Elevation.}

We saw how to transform a polynomial application into a linear mapping by 
adding extra variables representing the successive products and powers of every
variable. This information can be useful in order to generate invariants but in
fact, most of the time, this is not enough. In our running example of 
section~\ref{prel}, $g(x,y) = (x + y^2, y + 1)$, the degree of the mapping is 
$2$ but there exists no invariant of degree 2 for this loop. 
In order to deal with higher-degree invariants, 
we need not just to linearize $g$, 
we also have to add more variables to our study. 
As we can represent monomials of variables of a solvable mapping as linear
applications, we can extend the method to generate higher degree monomials such
as $y^3$ for example : we \emph{elevate} $g$ to a higher degree.
%
The process of elevation is described in~\cite{muller2004precise} as a way to
express polynomial relations on a linear program. 
%
%
\propinput{num_variables}
\paragraph{Note.} The complexity of the transformation is \emph{polynomial}
for $d$ or $n$ fixed. The lowerization algorithm can be used as shown above by 
adding variables computing the high degree monomials we want to linearize.
Moreover, {$n+d \choose d$} is an upper bound and in practice, we usually need much less 
variables.
For instance, in our running example, we don't need to consider $x.y^2$.
Indeed, if we tried to linearize this 
monomial, we would end up with $x.y^2 = x.y^2 + x.y + x + y^4 + 2y^3 + y^2$,
a polynomial of degree $4$. 
%
%
Detecting that a monomial $m$ is relevant or not can be done by computing the 
degree of its transformation.
For example, the assignment of $x$ is a degree $2$ polynomial, so $x^2$ 
associated transformation will be of degree $4$.
Here, there is actually only two interesting monomials of degree 3, which are $xy$ and 
$y^3$.
Though those variables will be useless for the linearized mapping, they are still 
easily computable: $y_3' = y_3 + 3.y_2 + 3.y + 1$ and $xy = xy + x + y_2 + y_3$.
This limits the necessary variables to only $6$ ($x,y,y_2,y_3,xy,\1$) instead 
of ${5\choose2}=10$. 
This upper bound in only reached for affine transformations 
when searching for polynomial invariants, as all possible monomials need to be 
treated. 

  \subsection{Invariant generation}\label{inv_gen}
   
The transformation described previously doesn't linearize a whole 
program, but only a loop. Polynomial assignments must be performed before the 
loop starts to initialize the new monomials. The method we present only 
focuses on the loop behavior itself, allowing any kind of operation 
outside of the loop.

\paragraph{Eigenspace.}

Loop invariants are logical formulas satisfied at every step of a loop.
We can characterize them with two criteria : they have to hold at the 
beginning of the loop (initialization criterion) and if they hold at one 
step, then they hold at the next step (heredity criterion).
%
%
Our technique is based on the discovery of linear combinations of variables 
that are equal to 0 and satisfying the heredity criterion.
For example, the loop of section~\ref{overview} admits the formula
$-6.x + y -3.y_2 + 2y_3 = k$ as a good invariant candidate.
Indeed, if we set $k$ in accordance with the values of the variables at the 
beginning of the loop, then this formula
will be true for any step of the loop.
We call such formulas \emph{semi-invariants}.

\begin{Def}\label{semi-inv}

 Let $\varphi : \mathbb{K}^n \mapsto \mathbb{K}$ and 
 $f :\mathbb{K}^n \mapsto \mathbb{K}^n$ two linear mappings.
 $\varphi$ is a semi-invariant for $f$ iff $\forall X$, 
 $\varphi(X) = 0 \Rightarrow \varphi(f(X)) = 0$.
 \end{Def}

\begin{Def} 
 Let $\varphi : \mathbb{K}^n \mapsto \mathbb{K}$, $f :\mathbb{K}^n \mapsto \mathbb{K}^n$ and $X \in \mathbb{K}^n$.
 $\varphi$ is an invariant for $f$  with initial state $X$ iff
$\varphi(X) = 0$
and
$\varphi$ is a semi-invariant for $f$.

\end{Def}
 
The key point of our technique relies on the fact that if there exists 
$\lambda, f^*(\varphi) = \lambda \varphi$, then we know that $\varphi$ 
is a semi-invariant. 
Indeed, we can rewrite definition~\ref{semi-inv} by 
$\scalprod{\varphi}{x} = 0 \Rightarrow \scalprod{\varphi}{f(x)} = 0$.
By linear algebra, we have
$\scalprod{\varphi}{f(x)} = \scalprod{f^*(\varphi)}{x}$, with 
$f^*$ the dual of $f$.
If $\exists\lambda, f^*(\varphi) = \lambda \varphi$, then we 
can deduce that 
$\scalprod{\varphi}{x} = 0 \Rightarrow \lambda\scalprod{\varphi}{x} = 0$.
This formula is always true, thus we know that $\varphi$ is a semi-invariant.
Such $\varphi$ are commonly called \emph{eigenvectors} of $f^*$. 
We will not adress the problem of computing the eigenvectors of an application
as this problem have been widely studied (in~\cite{pan1999complexity} for 
example).
 
Recall our running example $g(x,y) = (x + y^2,y+1)$, linearized by the
application
$f(x,y,y_2,xy,y_3,\1) = (x + y_2,y+\1,y_2 + 2y + \1, xy + x + y_2 + y_3,y_3 + 3y_2 + 3y + \1,\1)$.
$f^*$ admits $e_1 = (-6,1,-3,0,2,0)^t$ and $e_2 = (0,0,0,0,0,1)^t$ as 
eigenvectors associated to the eigenvalue $\lambda = 1$. 
It means that if $\scalprod{k_1.e_1 + k_2e_2}{x} = 0$, then 
\begin{eqnarray*} 
\scalprod{k_1.e_1 + k_2e_2}{f(X)} &=& \scalprod{f^*(k_1.e_1 + k_2e_2)}{X}\\
~ &=& \scalprod{\lambda(k_1.e_1 + k_2e_2)}{X} \\
~ &=& 0
\end{eqnarray*}
In other words, $\scalprod{k_1.e_1 + k_2e_2}{X} = 0$ is a semi-invariant. 
Then, by expanding it, we can find that $-6.x + y -3.y_2 + 2y_3 = k$, 
with $k = -\frac{k_2}{k_1}$ is a semi-invariant.
In terms of the original variables, we have thus $-6.x + y -3.y^2+ 2y^3 = k$.

Being an eigenvector of $f^*$ does not just guarantee
a formula to be a semi-invariant of a loop transformed by $f$.
This is also a necessary condition.

\propinput{invar_eigen}

It is now clear that the set of invariants is exactly the union of all 
eigenspaces of $f^*$, i.e. a vector space union (which is not a vectorial 
space itself).
An element $\varphi$ of $E_\lambda$ of basis $\{e_1,...e_n\}$
is a linear combination of $e_1,...,e_n$:
\[
\varphi = \sum\limits_\mathit{k=1}^n k_ie_i
\]
The parameters $k_i$ can be chosen with respect to the initial state of the 
loop.
%
%
%

\paragraph{Expression of eigenvectors as invariants.}


More than a syntactic sugar, the variable $\1$ brings interesting properties
over the kind of invariants we generate for an application $f$.
The vector $e_\1$ such that $\scalprod{e_\1}{X} = \1$ is always an eigenvector 
associated to the eigenvalue $1$. Indeed, by definition $f(\1)=\1$, hence
$f^*(e_\1)=e_\1$.
%
%
For example, let's take the mapping $f(x,y,xy,\1) = (2x, \frac{1}{2}y + 1,xy + 2x,\1)$.
This mapping admits 3 eigenvalues : $2,~\frac{1}{2}$ and $1$.
%
%
%
There exists two eigenvectors for the eigenvalue $1$ : $(-2,0,1,0)$ and
$(0,0,0,1) = e_\1$.
We have then the semi-invariant $k_1.(-2x + xy) + k_2 = 0$, or 
$-2x + xy = \frac{-k_2}{k_1}$.
This implies that the two parameters $k_1$ and $k_2$ can be reduced to only 
one paramter $k = \frac{-k_2}{k_1}$, 
which simplifies a lot the equation by providing a way to compute the parameter
at the initial state if we know it.
For our example, $\frac{-k_2}{k_1}$ would be $-2x_\mathit{init} + x_\mathit{init}.y_\mathit{init}$,
where $x_\mathit{init}$ and $y_\mathit{init}$ are the initial values of $x$ and $y$.
More generally, 
each eigenvector associated to $1$ gives us an invariant $\varphi$ 
that can be rewritten as $\varphi(X) = k$, where $k$ is inferred from the 
initial value of the loop variables. 
%

We can generalize this observation to eigenvectors associated to any
eigenvalue.
To illustrate this category, let us take as example $f(x,y,z) = (2x,2y,2z)$. 
%
Eigenvectors associated to $2$ are $e_1=(1,0,0)$, $e_2=(0,1,0)$ and $e_3=(0,0,1)$, thus
$k_1x + k_2y + k_3z= 0$
is a semi invariant, for any $k_1$, $k_2$ and $k_3$ satisfying the formula 
for the initial condition of the loop. 
However, if we try to set e.g. $k_1=k_2 = 1$, using
$x + y + kz = 0$ as semi invariant, we won't be able to find a proper invariant when
$y_\mathit{init}$ or $x_\mathit{init} \neq 0$ and $z_\mathit{init} = 0$.
Thus, in order to keep the genericity of our formulas, we cannot afford to 
simplify
the invariant as easily as we can do for invariants associated to the 
eigenvalue $1$. 
Namely for every $e_i$, we have to test whether $\scalprod{e_i}{X_\mathit{init}}=0$. 
For each $e_i$ for which this is the case, $\scalprod{e_i}{X} = 0$ is itself an 
invariant if $\scalprod{e_i}{X_\mathit{init}} = 0$.
However, if there exists an $i$ such that $\scalprod{e_i}{X_\mathit{init}} \neq 0$,
then we can simplify the problem. 
For example, we assume that $z_\mathit{init} \neq 0$. Then 
$k_1x_\mathit{init} + k_2y_\mathit{init} + k_3z_\mathit{init}= 0 \Leftrightarrow \frac{k_1x_\mathit{init} + k_2y_\mathit{init}}{z_\mathit{init}} =- k_3$.
We know then that $k_1x + k_2y =  \frac{k_1x_\mathit{init} + k_2y_\mathit{init}}{z_\mathit{init}}z$ 
is a semi-invariant.
By writing $g(k_1,k_2) = \frac{k_1x_\mathit{init} + k_2y_\mathit{init}}{z_\mathit{init}}$, we have 

$\left\{{
\begin{array}{lll}
 x &=& g(1,0)z\\
 y &=& g(0,1)z
\end{array}
}
\right.$

As $g$ is a linear application, these two invariants implies that
$\forall k_1,k_2, k_1x + k_2y = g(k_1,k_2)z$ is a semi-invariant.

\propinput{movable_constants}

We are now able to use pairs of eigenvectors to express invariants by knowing 
the initial condition.

\paragraph{Algorithm.}
%
As we are restricting our study to solvable loops, that we know can be 
replaced without loss of generality by linear loops, we assume the input of 
this algorithm is a family of linear mappings.
We can easily compose them via 
their matrix representation. We end up with a new matrix $A$.
Computing the dual of $A$ is computing the matrix $A^T$.
Then, eigenvectors of $A^T$ can be computed
by many algorithms in the linear algebra literature~\cite{pan1999complexity}. 
As the eigenvalue problem is known to be polynomial, our invariant generation 
algorithm is also polynomial.

  \subsection{Extension of the method}~\label{complex}
    Let $A \in \mathcal{M}_n(\mathbb{Q})$. In the general case, $A$ admits
  irrational and complex eigenvalues and eigenvectors, which end up 
  generating irrational or complex invariants.
  We cannot accept such representation for a further analysis of the input 
  program because of the future use of these invariants, by SMT solvers for 
  example which hardly deal with non-rational numbers.
  For example, let us take the function $f(x,y) = (y,2x)$. 
  This mapping admits two eigenvalues : $\lambda_x = \sqrt2$ and 
  $\lambda_y = -\sqrt2$. 
  In this example, the previous method would output the invariants 
  $k.(x + \sqrt2 y) = 0$ and $k'.(x-\sqrt2 y) = 0$. 
  With $x$ and $y$ integers or rationals, this would be possible iff 
  $k = k' = 0$. 
  However, by considering the variable $xy$ the invariant generation procedure 
  outputs the invariant $k.(xy) = 0$, which is possible if $x$ or $y$ equals 0.
  This raises the issue of finding a product of variables that
will give us a rational invariant.
  We aim to treat the problem at its source : the algebraic character of the 
  matrix eigenvalues. A value $x$ is algebraic in $\mathbb{Q}$ if there exists 
  a polynomial $P$ in $\mathbb{Q}[X]$ such that $P(x) = 0$. 
  Assuming we have a geometric relation between the complex eigenvalues 
  $\lambda_i$  (i.e. a product $q$ of eigenvalues that is rational), we will build 
  a monomial $m$ as a product of variables $x_i$ associated to $\lambda_i$
  such that the presence of this monomial induces the
  presence of a rational eigenvalue, namely $q$.
  Moreover, a rational eigenvalue of a 
  matrix is always associated to a rational eigenvector. 
  Indeed, the kernel of a rational matrix is always a $\mathbb{Q}$-vectorial 
  space.
%
  If $\lambda \in \mathbb{Q}$ is an eigenvalue of $A$, then 
  $A - \lambda.Id$ is a rational matrix and its kernel is not empty.

\begin{Def}

Let $A \in \mathcal{M}_n(\mathbb{Q})$ .
We denote $\Psi_d(A)$ the \emph{elevation} matrix such that 
$\forall X = (x_1,...,x_n)\in\mathbb{Q}^n, \Psi_d(A).p(X) = p(A.X)$, with
$p \in (\mathbb{Q}[X]^k)$ a polynomial associating $X$ to all possible 
monomials of degree $d$ or lower.
\end{Def}
For example, if we have  $A = \left(
  \begin{array}{cc}
    a & b\\
    c & d
  \end{array}\right)$ as a transformation 
  for $X = (x,y)$, \ifthenelse{\longarticle = 1}{and $x \prec y$,}{} we have
as transformation for the variables $(x^2,xy,y^2,x,y)$ the matrix 
  \[\Psi_2(A) = \left(\begin{array}{ccccc}
    a^2  & 2ab& b^2 & 0 & 0\\
    ac  & ad + bc& bd & 0 & 0\\
    c^2 & 2cd & d^2 & 0 & 0\\
    0 & 0 & 0 & a & b\\
    0 & 0 & 0 & c & d
  \end{array}\right)\]

\propinput{prod_eigen}
We can generalize this property for more variables. After working with two variables,
we get a new matrix with new variables that we can combine similarly, and so on.
Thanks to this property, if we have a multiplicative relation between 
eigenvalues we are able to create \emph{home-made} variables in the elevated 
application whose presence implies the presence of rational eigenvalues.

  Though we could brute-force the search of rational products of irrational 
  eigenvalues in order to find all possibilities of variable products that have
  rational eigenvalues, we could search for algebraic
  relations, i.e. multiplicative relations between 
  algebraic values. 
  This subject is treated in~\cite{kauers2008computing} and
  we will not focus on it. 
  However, we can guarantee that there exists at least one monomial having a 
  rational eigenvalue.
  Indeed, it is known that the product of all eigenvalues of a rational matrix 
  is equal to its determinant.
  As the determinant of a rational matrix is always rational, we know that the
  product of all variables infers the presence of the determinant of the matrix
  as eigenvalue of the elevated matrix.
  Coming back to 
  the previous example, we have the algebraic relation 
  $\lambda_x.\lambda_y = -2$. If we consider the evolution of $xy$, we have 
  $(xy') = 2xy$. Note that the eigenvalue associated to 
  $xy$ is $2$ and not $-2$. Indeed, we know that $A = P^{-1}JP$, with 
  
  $P =  \left(\begin{array}{cc}
    1 & -1\\
    \sqrt2 & -\sqrt2 
  \end{array}\right)$
  
  and $J$ an upper-triangular matrix, which means the eigenvalues of $A$ are
  on the diagonal of $J$.
  $xy$ in the base of $J$ would be 
  $(x + \sqrt2y)(x - \sqrt2y) = x^2 - 2y^2$, and we have well
  $\lambda_x^2 - 2\lambda_y^2 = -2$.
 
 Finally, by knowing that $\lambda_x^2 = 2$, $\lambda_y^2 = 2$ and 
 $\lambda_x\lambda_y = -2$, we will consider the variables $x^2$, $y^2$
 and $xy$ in our analysis of $f$.
 We can deduce new semi-invariants from these variables :
 $k_1(xy) + k_2(2x^2 + y^2) = 0$ with the eigenvectors associated to $2$ and 
 $k.(y^2 - 2x^2) = 0$ with the eigenvector associated to $-2$.

  \subsection{Multiple loops}~\label{mult_loop}
  In this short section, we present our method to treat non-deterministic loops,
i.e. loops with non-deterministic conditions. 
At the beginning of each iteration, the loop 
can choose randomly between all its bodies.
This representation is equivalent to the definition in section~\ref{prel}.

\begin{Def}
 Let $F = \{A_i\}_{1 \leqslant i \leqslant n}$ a family of matrices and $Inv(F)$ 
 the set of invariants of a loop whose different bodies can be encoded by 
 elements of $F$. 
 
 $Inv(F) = \{\varphi | \forall X, \varphi.X = 0 \Rightarrow \bigwedge\limits_{i=1}^n \varphi.A_i.X = 0 \}$
\end{Def}

\propinput{invar_inter}

As the set of invariants of a single-body loop are a vectorial 
spaces union, 
its intersection with another set of invariants is also a 
vector space union.
Although we do not consider the condition used by the program to choose
the correct body, we still can discover useful invariants. 
Let us consider the following example, taken 
from~\cite{rodriguez2007generating}, that computes the product of x and y in 
variable z :

\begin{lstlisting}
 while (*) do
    (x,y,z) := (2x, (y-1)/2, x + z)
    OR   
    (x,y,z) := (2x, y/2, z)
 done
\end{lstlisting}

We have to deal with two applications : $f_1(x,y,z) = (2x, (y-1)/2, x + z)$ and
$f_2(x,y,z) = (2x, y/2, z)$.
The elevation to the degree $2$ of $f_1$ and $f_2$ returns applications having 
both 10 eigenvectors. 
For simplicity, we focus on invariants associated to the eigenvalue 
$1$.
\begin{multicols}{2}

$f_1^*$ has $4$ eigenvectors $\{e_i\}_{i\in [1,4]}$ associated to 1 such that

\begin{itemize}
 \item $\scalprod{e_1}{X} = -x + xy$
 \item $\scalprod{e_2}{X} = x + z$
 \item $\scalprod{e_3}{X} = xz + x^2 + z^2$
 \item $\scalprod{e_4}{X} = \1$
\end{itemize}


$f_2^*$ also has $4$ eigenvectors $\{e'_i\}_{i\in [1,4]}$ associated to 1 such that 

\begin{itemize}
\item $\scalprod{e'_1}{X} = xy$
\item $\scalprod{e'_2}{X} = z$
\item $\scalprod{e'_3}{X} = z^2$ 
\item $\scalprod{e'_4}{X} = \1$
\end{itemize}

\end{multicols}

First, we notice that $e_4 = e'_4$.
Then, we can see that $\scalprod{e_1 + e_2}{X} = xy + z = \scalprod{e'_1 + e'_2}{X}$.
Thus, $e_1 + e_2 = e'_1 + e'_2$.
Eventually, we find that 
$e_1 + e_2 + k.e_4 \in (Vect(\{e_i\}_{i\in [1,4]}) \cap Vect(\{e'_i\}_{i\in [1,4]}))$.
That's why $(\scalprod{e_1 + e_2 + k.e_4}{X} = 0)$ is a semi-invariant for 
both $f_1$ and $f_2$, hence for the whole loop.
Replacing $\scalprod{k.e_4}{X}$ by $k = -k'$ and $\scalprod{e_1 + e_2}{X}$ by $xy + z$ gives us $xy + z = k'$.
\paragraph{Algorithm.}
The intersection of two vector spaces corresponds to the vectors that both 
vector spaces have in common. 
It means that such elements can be expressed by elements of the base of each
vector space.
Let $B_1$ and $B_2$ the bases of the two vector spaces.
If $e \in \vect\{B_1\}$ and $e \in \vect\{B_2\}$, then $e \in \ker\{(B_1B_2)\}$.
To compute the intersection of a vector space union, we just have to
compute the kernels of each combination of vector space in the union.

 \section{Implementation and experimentation}\label{results}
 
    In order to test our method, we implemented an invariant generator as a
    plugin of Frama-C~\cite{Kirchner2015}, a framework for the verification of 
    C programs written in OCaml.
    Tests have been made on a Dell Precision M4800 with 16GB RAM and 8 cores. Time
    does not include parsing time of the code, but only the invariant 
    computation from the Frama-C representation of the program to the formulas.
    Moreover, our tool doesn't implement the extension of our method and may 
    output irrational invariants or fail on complex eigenvalues. 
    Benchmark is available at~\cite{poly_bench}.
   The second column of the table~\ref{table:results} represents the number of variables used
   in the program. The third column represents the invariant degree used for 
   \toolname and \fastind. 
   The last three columns are the computation time of the tools in \emph{ms}.
   O.O.T. represents an aborted 
   ten minutes computation and -- indicates that no invariant is found.

\begin{table}[htbp]
 \caption{Performance results with our implementation \toolname}\label{table:results}
  \begin{center}
     \begin{tabular}{|c||c|c||r|r||r|}
      \multicolumn{1}{c}{} & \multicolumn{1}{c}{Program}&  & \multicolumn{1}{c}{} & \multicolumn{1}{c}{Time~(in ms)} & \multicolumn{1}{c}{}\\
	\hline
	\hline
	 & & & & &\\
	Name & ~Var~ & ~Degree~ & \multicolumn{1}{c|}{\aligator~\cite{kovacs2008aligator}~} & \multicolumn{1}{c||}{~\fastind~\cite{cachera2014inference}} &\multicolumn{1}{c|}{~\toolname~~}\\
	 & & & & &\\
	\hline
	\hline
	divbin& 5 &2 & 80 & 6 & 2.5 \\
	\hline
	hard& 6 & 2 & 89 & 13 & 2 \\
	\hline
	mannadiv & 5 & 2 & 27 & 6 & 2 \\
	\hline
	sqrt& 4 & 2 & 33 & 5 & 1.5 \\
	\hline 
	djikstra& 5 & 2 & 279 & 31 & 4 \\
	\hline
	euclidex2 & 8 & 2 & 1759 & 10 & 6 \\
	\hline
	lcm2& 6 & 2 & 175 & 6 & 3 \\
	\hline
	prodbin& 5 & 2 & 100 & 6 & 2.5 \\
	\hline
	prod4 & 6 & 2 &13900& -- & 8 \\
	\hline
	fermat2 & 5 & 2 & 30 & 9 & 2 \\
	\hline
	knuth & 9 & 3 & \multicolumn{1}{c|}{~O.O.T.} & 347 & 192 \\
	\hline
	eucli\_div & 3 & 2 & 13 & 6 & 2\\
	\hline
	cohencu & 5 & 2 & 90 & 5 & 2\\
	\hline
	read\_writ & 6 & 2 & 82 & -- & 12\\
	\hline
	illinois & 4 & 2 & \multicolumn{1}{c|}{~O.O.T.} & -- & 8\\
	\hline
	mesi & 4 & 2 & 620 & -- & 4\\
	\hline
	moesi & 5 & 2 & \multicolumn{1}{c|}{~O.O.T.} & -- & 8\\
	\hline
	petter\_4 & 2 & 10 & 19000 & 37 & 3 \\
	\hline
	petter\_5 & 2 & 10 & \multicolumn{1}{c|}{~O.O.T.} & 37 & 2\\
	\hline 
	petter\_6 & 2 & 10 & \multicolumn{1}{c|}{~O.O.T.} & 37 & 2\\
	\hline
      \end{tabular}
      
   \end{center}
      \vspace{2ex}

\end{table}
    
All the tested functions are examples for which the presence of a polynomial
invariant is compulsory for their verification.
The choice of high degree for some functions is 
motivated by our will to show the efficiency of our tool to find high degree invariants
as choosing a higher degree induces computing a bigger set of relations. 
%
%
    In the other cases, degree is choosen for its usefulness.

    For example in figure~\ref{fig:pig_test} we were interested in finding the 
    invariant $x + qy = k$ for eucli\_div. 
    That's why we set the degree to $2$.
%
\begin{figure}[htbp]
  \begin{framed}
  \begin{minipage}{0.4\textwidth}
    
    \begin{center}
      \vspace{-5ex}
      Input : degree = 2
    \end{center} 
    \begin{lstlisting}[language=C]
int eucli_div(int x, int y){
  int q = 0;
  while (x > y) {
    x = x-y;
    q ++;
  }
  return q;
}
    \end{lstlisting}
  \end{minipage}
   \hfill
   \begin{minipage}{0.57\textwidth}
     \begin{center}
	Frama-C output : 
     \end{center}
     \begin{lstlisting}[language=C]
      int eucli_div(int x, int y){
        int q = 0;
        int k = x + y*q;
        // invariant x + y*q = k;
        while (x > y) {
	  x = x-y;
	  q ++;
        }
        return q;
      }
     \end{lstlisting}
      \end{minipage}

  \end{framed}

     \caption{\label{fig:pig_test} Euclidean division C loop and generation 
     of its associated invariants.}
    \end{figure}
Let $X$ be the vector of variables $(x,y,q,xq,xy,qy,y_2,x_2,q_2,\1)$. 
The matrix $A$ representing the loop in figure~\ref{fig:pig_test} has only one 
eigenvalue : $1$.
There exist 4 eigenvectors $\{e_i\}_{i\in[1;4]}$ associated to $1$ in $A$,
so $\scalprod{\sum\limits_{i = 1}^4k_ie_i}{X} = 0$ is a semi-invariant.
One of these eigenvectors, let's say $e_1$, correspond to the constant 
variable, i.e. $e_1.X = \1 = 1$, thus we have $\scalprod{\sum\limits_{i = 2}^4k_ie_i}{X} = -k_1$ 
as invariant.
In our case, $\scalprod{e_2}{X} = y$, $\scalprod{e_3}{X} = x + yq$ and $\scalprod{e_4}{X} = y_2$.
We can remove $(y = k)$ and $(y_2 = k)$ that are evident because $y$ does
not change inside the loop.
The remaining invariant is $x + yq = k$.

 \section{Conclusion and future work}

    We presented a simple and effective method to generate non-trivial 
    invariants. 
    One of its great advantages is to only rely on linear algebra theory, and 
    generate modular invariants. 
    Still our method has some issues that we are currently investigating.
    First, it is incomplete for integers : invariants we generate are 
    only correct for rationals. 
    Perhaps surprisingly, this issue does not come from the invariant 
    generation, but 
    from the linearization procedure which badly takes into account the 
    division.
    For example in C, the operation $x' = \frac{x}{2}$ with $x$ uneven returns 
    $\frac{x - 1}{2}$. 
    This behavior is not taken into account by the elevation, which can freely
    multiply this $x$ by a variable $y$ with $y' = 2y$.
    This returns the assignment $xy' = xy$ which is false if $x$ is odd.
    Next, we do not treat interleaving loops as we cannot yet compose 
    invariants with our generation technique.
    The tool has been successfully implemented as an independent tool of Frama-C.

    Our next step is to use those invariants 
    with the Frama-C tools Value (a static value 
    analyser) and WP (a weakest precondition calculus API) to apply a 
    CEGAR-loop on counter-examples generated by CaFE, a temporal logic model 
    checker based on~\cite{alur2004temporal}.
    Also, we want the next version of the tool to handle irrational 
    eigenvalues as decribed in section~\ref{complex}.

\bibliography{main}

\begin{thebibliography}{10}

\bibitem{alur2004temporal}
R.~Alur, K.~Etessami, and P.~Madhusudan.
\newblock A temporal logic of nested calls and returns.
\newblock In {\em {TACAS} 2004}, pages 467--481, 2004.

\bibitem{basu1975proving}
S.~K. Basu and J.~Misra.
\newblock Proving loop programs.
\newblock {\em {IEEE} Trans. Software Eng.}, 1(1):76--86, 1975.

\bibitem{beyer2007path}
D.~Beyer, T.~A. Henzinger, R.~Majumdar, and A.~Rybalchenko.
\newblock Path invariants.
\newblock In {\em {ACM} {SIGPLAN} 2007 Conference on Programming Language
  Design and Implementation}, pages 300--309, 2007.

\bibitem{botella2009automating}
B.~Botella, M.~Delahaye, S.~H.~T. Ha, N.~Kosmatov, P.~Mouy, M.~Roger, and
  N.~Williams.
\newblock Automating structural testing of {C} programs: Experience with
  {PathCrawler}.
\newblock In {\em 4th International Workshop on Automation of Software Test,
  {AST} 2009}, pages 70--78, 2009.

\bibitem{cachera2014inference}
D.~Cachera, T.~P. Jensen, A.~Jobin, and F.~Kirchner.
\newblock Inference of polynomial invariants for imperative programs: {A}
  farewell to {G}r{\"{o}}bner bases.
\newblock {\em Sci. Comput. Program.}, 93:89--109, 2014.

\bibitem{poly_bench}
E.~Carbonell.
\newblock Polynomial invariant generation.
\newblock
  \url{http://www.cs.upc.edu/~erodri/webpage/polynomial_invariants/list.html}.

\bibitem{clarke2000counterexample}
E.~M. Clarke, O.~Grumberg, S.~Jha, Y.~Lu, and H.~Veith.
\newblock Counterexample-guided abstraction refinement.
\newblock In {\em {CAV} 2000}, pages 154--169, 2000.

\bibitem{cooper2001operator}
K.~D. Cooper, L.~T. Simpson, and C.~A. Vick.
\newblock Operator strength reduction.
\newblock {\em {ACM} Trans. Program. Lang. Syst.}, 23(5):603--625, 2001.

\bibitem{hoare1969axiomatic}
C.~A.~R. Hoare.
\newblock An axiomatic basis for computer programming.
\newblock {\em Commun. {ACM}}, 12(10):576--580, 1969.

\bibitem{hoare2003verifying}
C.~A.~R. Hoare.
\newblock The verifying compiler: {A} grand challenge for computing research.
\newblock {\em J. {ACM}}, 50(1):63--69, 2003.

\bibitem{hojjat2012accelerating}
H.~Hojjat, R.~Iosif, F.~Konecn{\'{y}}, V.~Kuncak, and P.~R{\"{u}}mmer.
\newblock Accelerating interpolants.
\newblock In {\em {ATVA} 2012}, pages 187--202, 2012.

\bibitem{johnson1975finding}
D.~B. Johnson.
\newblock Finding all the elementary circuits of a directed graph.
\newblock {\em SIAM Journal on Computing}, 4(1), 1975.

\bibitem{kauers2008computing}
M.~Kauers and B.~Zimmermann.
\newblock Computing the algebraic relations of {C}-finite sequences and
  multisequences.
\newblock {\em J. Symb. Comput.}, 43(11):787--803, 2008.

\bibitem{Kirchner2015}
F.~Kirchner, N.~Kosmatov, V.~Prevosto, J.~Signoles, and B.~Yakobowski.
\newblock {Frama-C: A software analysis perspective}.
\newblock {\em Formal Aspects of Computing}, 27(3), 2015.

\bibitem{kovacs2008aligator}
L.~Kov{\'{a}}cs.
\newblock Aligator: {A} mathematica package for invariant generation (system
  description).
\newblock In {\em Automated Reasoning, 4th International Joint Conference,
  {IJCAR} 2008}, pages 275--282, 2008.

\bibitem{kovacs2010complete}
L.~Kov{\'{a}}cs.
\newblock A complete invariant generation approach for {P}-solvable loops.
\newblock In {\em Perspectives of Systems Informatics, 7th International Andrei
  Ershov Memorial Conference, {PSI} 2009}, pages 242--256, 2009.

\bibitem{mayr1989membership}
E.~Mayr.
\newblock {\em Membership in polynomial ideals over Q is exponential space
  complete}.
\newblock Springer, 1989.

\bibitem{muller2002polynomial}
M.~M{\"{u}}ller{-}Olm and H.~Seidl.
\newblock Polynomial constants are decidable.
\newblock In {\em Static Analysis, 9th International Symposium, {SAS} 2002},
  pages 4--19, 2002.

\bibitem{muller2004precise}
M.~M{\"{u}}ller{-}Olm and H.~Seidl.
\newblock Precise interprocedural analysis through linear algebra.
\newblock In {\em {POPL} 2004}, pages 330--341, 2004.

\bibitem{pan1999complexity}
V.~Y. Pan and Z.~Q. Chen.
\newblock The complexity of the matrix eigenproblem.
\newblock In {\em Proceedings of the Thirty-First Annual {ACM} Symposium on
  Theory of Computing, May 1-4, 1999, Atlanta, Georgia, {USA}}, pages 507--516,
  1999.

\bibitem{rodriguez2007generating}
E.~Rodr{\'{\i}}guez{-}Carbonell and D.~Kapur.
\newblock Generating all polynomial invariants in simple loops.
\newblock {\em J. Symb. Comput.}, 42(4):443--476, 2007.

\bibitem{tarjan1972depth}
R.~E. Tarjan.
\newblock Depth-first search and linear graph algorithms.
\newblock {\em SIAM journal on computing}, 1(2), 1972.

\end{thebibliography}
\bibliographystyle{abbrv}

  \ifthenelse{\longarticle = 1 \and \annexedProof = 1}
    {
    \newpage
    \section{Appendix}
	\renewcommand{\inAnnex}{1}
	\setcounter{thm}{0}
	\setcounter{prop}{0}

  \paragraph{Linearization theorem.}
  
  The following justifies the linearization theorem of section~\ref{poly2lin}
  \propinput{linearizable}
  
  \paragraph{Complexity.}
  
  The following justifies the complexity theorem of elevation in section~\ref{poly2lin}
  \propinput{num_variables}
  
  \paragraph{Invariants set.}
  
  The following justifies the completude of the invariant generation procedure 
  in section~\ref{inv_gen}
  \propinput{invar_eigen}
  
  \paragraph{Constant deplacement.}
  The following justifies the operation of preserving the 
  equivalency of two invariants by moving constants in section~\ref{inv_gen}.
  \propinput{movable_constants}
  
  \paragraph{Extension of the method.}
  
  The following justifies the theorema of construction of rational eigenvalue 
  in section~\ref{complex}.
  \propinput{prod_eigen}
  
  \paragraph{Multiple loop.}
  
  The following justifies the multiple loop proposition of section~\ref{complex}
  \propinput{invar_inter}

    }

\end{document}